\documentclass{article}
\usepackage{spconf,amsmath,graphicx}
\usepackage{algorithm}
\usepackage{algorithmic}
\usepackage{booktabs}
\usepackage{microtype}
\usepackage{graphicx}
\usepackage{hyperref}
\usepackage{multirow}
\usepackage{color}
\usepackage{amsthm,amsmath,amssymb}
\usepackage{mathrsfs}
\usepackage[caption=false,font=footnotesize]{subfig}

\title{End-to-end Silent Speech Recognition with Acoustic Sensing}

\name{Jian Luo, Jianzong Wang*\thanks{*Corresponding author: Jianzong Wang, jzwang@188.com}, Ning Cheng, Guilin Jiang, Jing Xiao}
\address{Ping An Technology (Shenzhen) Co., Ltd.}

\begin{document}

\maketitle
\begin{abstract}
Silent speech interfaces (SSI) has been an exciting area of recent interest. In this paper, we present a non-invasive silent speech interface that uses inaudible acoustic signals to capture people's lip movements when they speak. We exploit the speaker and microphone of the smartphone to emit signals and listen to their reflections, respectively. The extracted phase features of these reflections are fed into the deep learning networks to recognize speech. And we also propose an end-to-end recognition framework, which combines the CNN and attention-based encoder-decoder network. Evaluation results on a limited vocabulary (54 sentences) yield word error rates of 8.4\% in speaker-independent and environment-independent settings, and 8.1\% for unseen sentence testing. 
\end{abstract}


\section{Introduction}
\label{sec:intro}

With the rapid development of speech recognition and natural language processing, voice user interface (VUI) has become a fundamental use case for today's smart devices (e.g., smartphone, smartwatch, laptop, smart speaker, and smart appliance). However, voice interaction suffers from several limitations that severely hinder its usage in daily life. First, audible speech is not suitable in some scenarios, such as in a meeting or when someone is sleeping. Second, environmental noise, like traffic noise, industrial machinery noise, and speech from bystanders, can make speech recognition challenging or even impossible. Third, people are unlikely to use voice input in public areas due to its risk of privacy leakage.

Recent advances in silent speech recognition have opened up new possibilities to counterbalance the above limitations. Some methods are based on computer vision technology \cite{zhou2014review} to capture the visual features of lip movements. The adoption of deep learning substantially boosts the precision of vision-based speech recognition \cite{assael2016lipnet,chung2017lip}. However, these methods are highly sensitive to lighting conditions, which means they cannot work in dark environments. Some other works exploit a variety of face-worn sensors for speech sensing, such as EMG electrodes \cite{wand2016deep,wand2018domain,kapur2018alterego}, RFID tags \cite{wang2019rfid}, bone-conduction vibration sensors \cite{maruri2018v}. While a significant drawback of these works is that the skin-attached sensors are very invasive to users. Besides, the reattachment of the sensors may cause changes in the recorded signal, which will significantly degrade their performance \cite{wand2018domain}.

\begin{figure}{}
	\centering
	\includegraphics[width=3in]{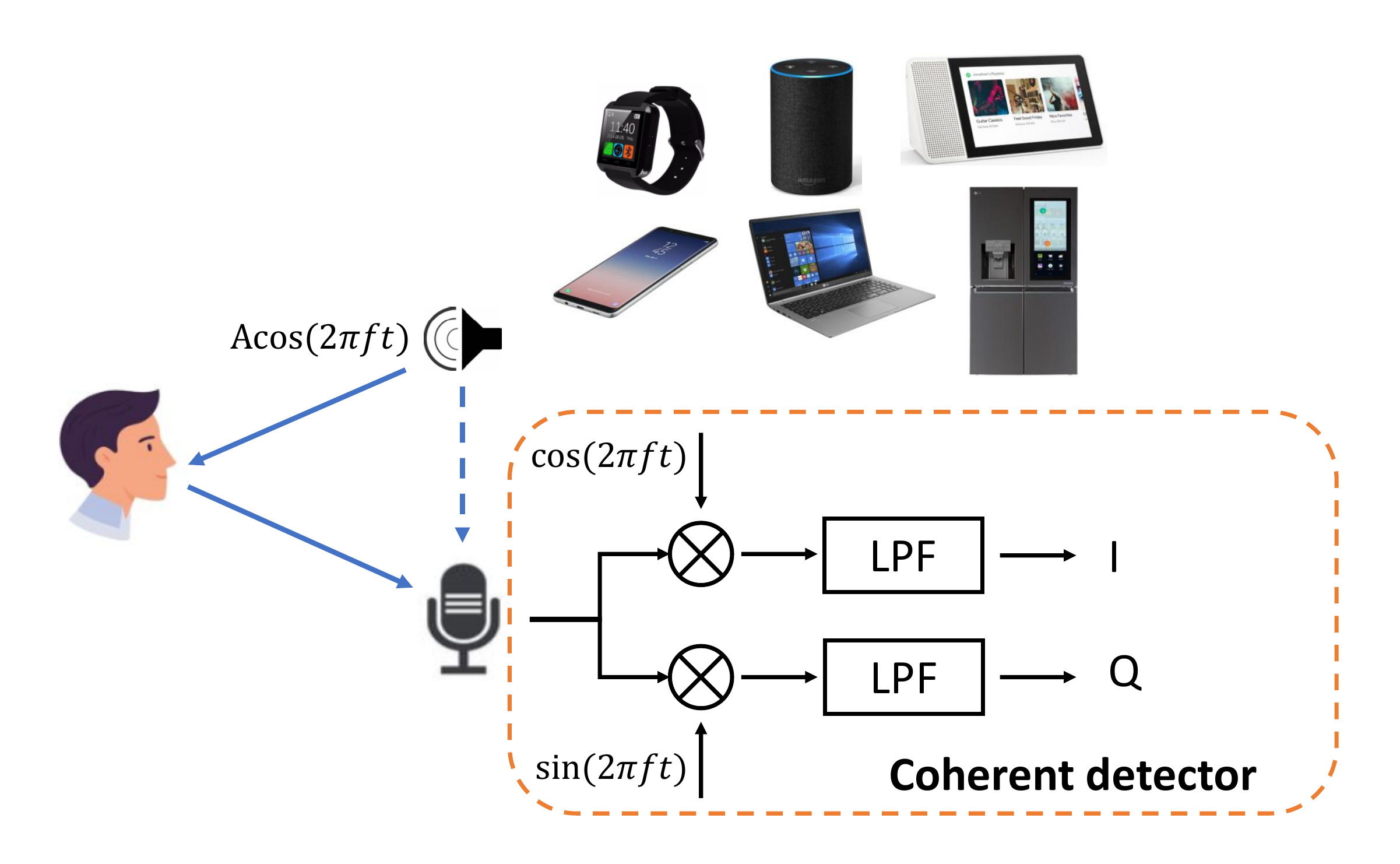}
	\caption{Sensing lip movements with acoustic signals generated by smart devices.}
	\label{lipreading2}
\end{figure} 
{}
Instead of the aforementioned approaches, another trend of works utilizes ultrasound. The use of high-frequency ultrasound (MHz level) has a long history in medical voice interface research \cite{Sonies1981}, aiming to provide an alternative to electrolarynx for some patients who have lost their voicebox. In recent research, most of the works use ultrasound to build real-time 2D tongue images \cite{denby2004,hueber2011,Hakoun2016}. However, the method using high-frequency ultrasound requires a special ultrasonic-imaging device that is not convenient for daily uses. Therefore, researchers also developed some applications based on low-frequency (LF) ultrasound. Inspired by gesture recognition \cite{gupta2012soundwave,ruan2016audiogest}, they employed LF ultrasound to detect lip movements \cite{zhang2017hearing,tan2017silenttalk,lu2019lip}, instead of creating tongue images. In these works, acoustic sensing systems for simple lip-reading mainly use the Doppler shift of the received signal. Nevertheless, due to limited frequency precision, the Doppler shift can only provide coarse-grained estimation\cite{wang2016device}, which is not suitable to capture subtle lip movements.

In this paper, we put forward a non-invasive silent speech interface, using LF ultrasound with some critical modifications. Our contributions focus on the followings:

(1). Propose an end-to-end silent speech interface for continuous recognition using acoustic signals, which is completely non-invasive and needs no extra hardware modification except existing smart devices.

(2). Leverage the phase information of the received signals, instead of Doppler shift, to obtain the fine-grained estimation of lip movements, and carefully design the signal preprocessing pipeline.

(3). Employ CNNs to extract representative features, and use the attention-based encoder-decoder network to enable end-to-end recognition as well as learn the underlying language model.  

Our method can be deployed on existing smart devices, exploiting speaker and microphone for lip-reading. People don't need to wear any sensors, but only need to move the devices near their mouths. As Figure~\ref{lipreading2} depicts, when people speak to the devices, our system leverages the speaker to emit the inaudible signal and the microphone to listen to the signal reflected by moving lips. Then the system analyses the reflected signal to recognize speech.

\begin{figure*}
		\centering
		\subfloat[The Doppler shift]{\includegraphics[width=2.3in]{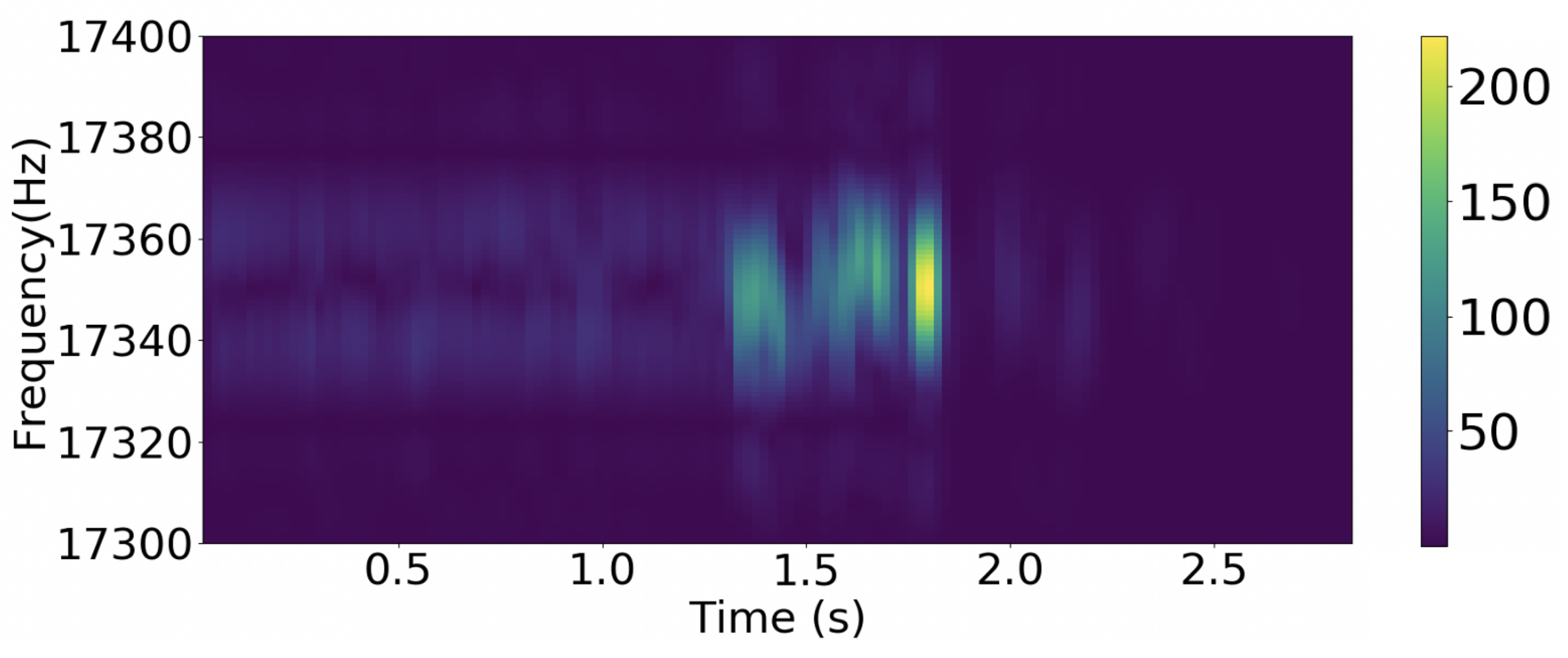}}
		\hspace{10pt}
		\subfloat[Phase]{\includegraphics[width=2.1in]{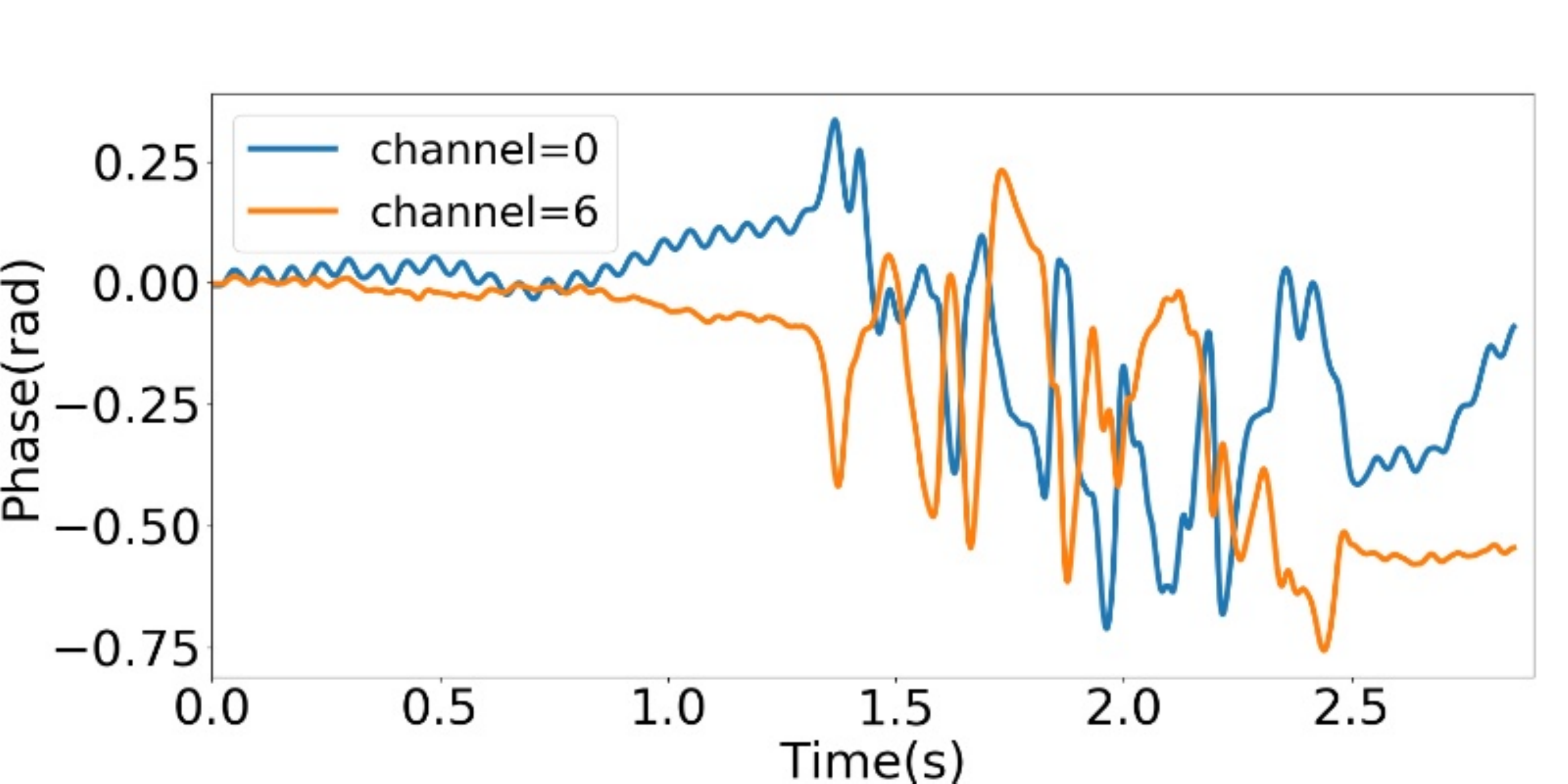}}
		\hspace{10pt}
		\subfloat[Phase delta]{\includegraphics[width=2.1in]{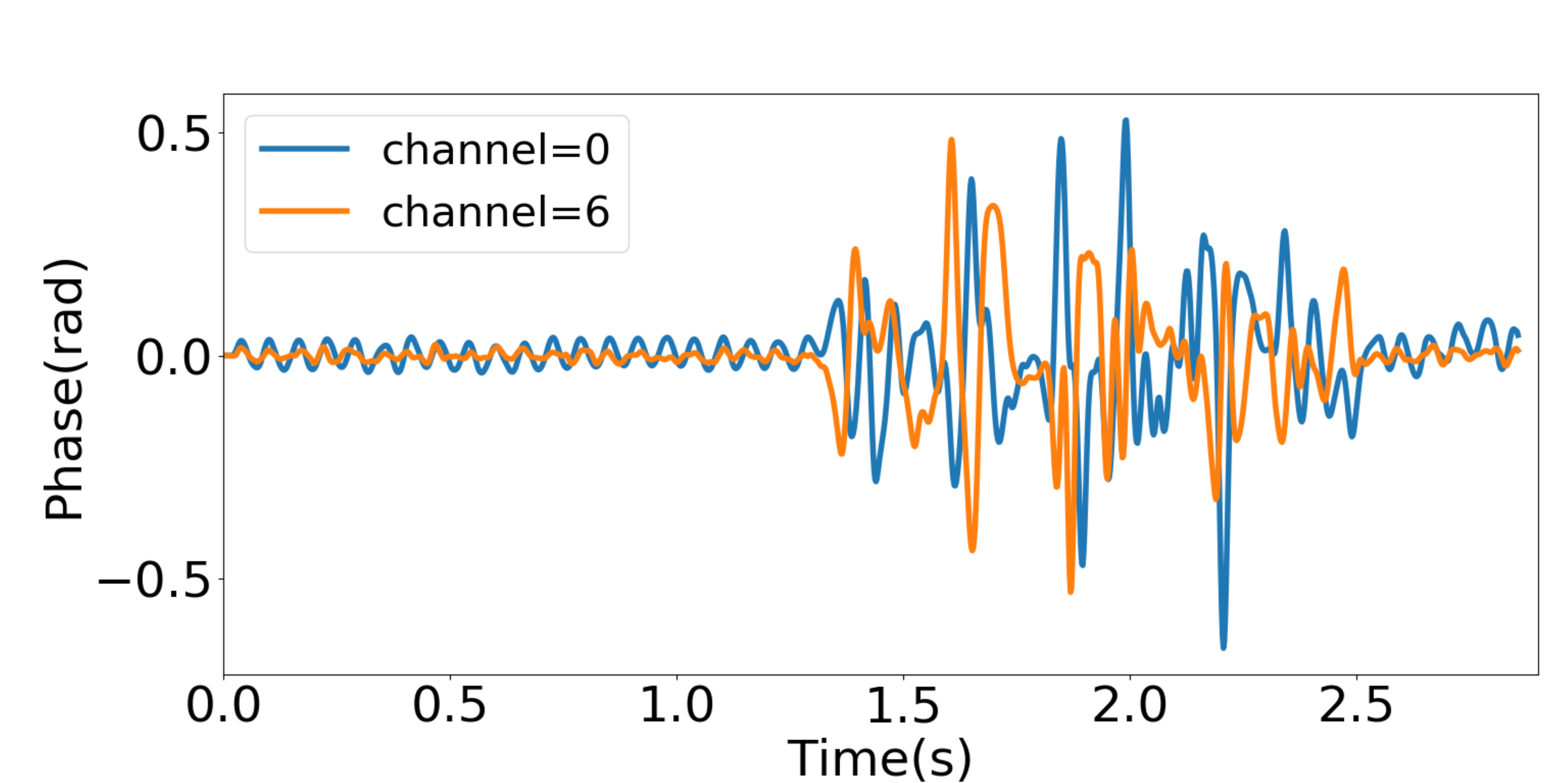}}
		\caption{Different acoustic signals of a lip movement from 1.2s to 2.5s when speaking the word ``WiFi". (a) shows the STFT result of the Doppler shift, where we can hardly observe signals after 1.8s. (b) and (c) shows the phase and phase delta signals, respectively. We demonstrate two frequency channels with $k=0,6$. It is clear that the phase delta performs the best.}
		\label{doppler_pham}
\end{figure*}  

\section{Signal Measurement and Preprocessing}

\subsection{Limitations of Doppler Shift}
Most existing smart devices can emit and record sound waves with frequency up to 23 kHz, and researchers showed that sound waves higher than 17 kHz are usually inaudible to most people\cite{rodriguez2014extended}. Therefore, the speaker and microphone of the devices can act as an active sonar to sense surroundings. Many researchers used the Doppler shift calculated by STFT to estimate movements. However, the resolution of STFT is limited by the fundamental constraints of time-frequency analysis. Figure~\ref{doppler_pham}(a) shows the STFT results of a moving lip from 1.2s to 2.5s when speaking the word ``WiFi". The frequency of the emitted signal is 17.35 kHz. We remove the inference of the Line-of-Sight (LOS) signal by calculating the difference between two successive samples\cite{lu2019lip} in the frequency-domain. Since small frequency variations are buried in the wide frequency band around 17.35 kHz, we can hardly observe the Doppler shift signals after 1.8s.

\subsection{Phase Measurement}  

To overcome the limitations of the Doppler shift, we leverage phase information of received signal to profile fine-grained lip movements. The wavelength of sound waves up to 17 kHz is less than 2 cm, meaning that a small movement of frequency will significantly change the phase of the received sound wave. Therefore, the signal phase is susceptible to subtle changes of propagation distance.

The phase signals can be calculated through the coherent detector, as Figure~\ref{lipreading2} depicts. Firstly, the inaudible signals reflected by moving lips are collected by the microphone of smartphone. Secondly, these signals are fed into low-pass filters (LPF) to get In-phase ($I$) component and Quadrature ($Q$) component respectively. Thirdly, these two components are combined together to get the phase features. Specificly, the speaker of smartphone emit the Continuous Wave (CW) signal of $Acos(2\pi ft)$, where $A$ is the amplitude, $t$ is the sampling index on time axis, and $f$ is the frequency of the sound, which is higher than 17 kHz. The sampling rate is 48 kHz. Without loss of generality, we assume there is only one propagation path $d_p(t)$, here $p$ denote propagation. And thus the received signal of reflection can be denoted as:
\begin{equation}
\begin{aligned}
R_p &= A_p cos(2\pi f(t-d_p(t)/c) - \theta_p)
\\
&= A_p cos(2\pi ft-\phi_p),
\end{aligned}
\end{equation}
where $c$ is the speed of the sound, and $\theta_p$ is the phase shift brought by the hardware. $A_p$ and $\phi_p$ are amplitude and phase, respectively. The received signal will be multiplied by $cos(2\pi ft)$ :
\begin{equation}
\begin{aligned}
R_p\times cos(2\pi ft) &= A_p cos(2\pi ft-\phi_p)\times cos(2\pi ft)
\\
&= \frac{A_p}{2} cos(4\pi ft-\phi_p) + \frac{A_p}{2} cos(\phi_p).
\end{aligned}
\end{equation}

The first term in the equation has a high frequency of $2f$ and thus can be removed by a LPF. Then, we can get the $I$ component of the base-band signal as $I_p = \frac{A_p}{2} cos(\phi_p)$. To reduce computational complexity, the $I$ component will pass through a moving average filter with a window size of 200 and an overlap of 0.5. This makes the sampling rate decreased from 48 kHz to 480 Hz.  Similarly, we can get the $Q$ component as  $Q_p = \frac{A_p}{2} sin(\phi_p)$. Then, these two components are combined as the real and imaginary parts of a complex signal:
\begin{equation}
B_p=\frac{A_p}{2}e^{-j\phi_p}.
\label{base-band}
\end{equation}

We can easily get the phase signals $\phi_p(t)$ from Equation~\ref{base-band}. Figure~\ref{doppler_pham}(b) shows the phase profile obtained from the same sound record that produces the spectrogram in Figure~\ref{doppler_pham}(a). We can clearly observe patterns caused by lip movements. The profiles exhibit significant fluctuations from 1.2s to 2.5s.

\subsection{Signal Preprocessing}

\textbf{Multi-frequency acoustic signals.} Wireless signals with different frequencies will experience different multipath fading when propagating in the air \cite{tse2005fundamentals}. Therefore, we simultaneously transmit sound waves at multiple frequencies to mitigate frequency selective fading as well as enhance the capability to profile multipath environments. In particular, we generate signal $A\sum_k cos[2\pi (f+k\delta f)t]$, which is the superposition of multi-frequency sound waves. $k$ depicts the $k$th frequency channel, and $\delta f$ is the frequency interval between adjacent channels. In the receiver, we get the phase values for each frequency using the corresponding coherent detector. All the frequencies fall into the band of 17$\sim$23 kHz. Considering the signal energy for each frequency and limited bandwidth, we set the number of channels $k$ to 8 and $\delta f$ to 700 Hz. 

\textbf{Multipath elimination.} The received signals are the mixtures of multipath signals. Besides dynamic signals caused by moving lips, there exist static signals including the LOS signal (i.e., the signal directly propagated from the speaker to microphone) and surrounding reflections (from face and body), which are usually much stronger than dynamic signals. Moreover, static signals may also change slowly with the movements of the face or body. Figure~\ref{doppler_pham}(b) shows that the phase still increases slowly after 2.5s (the end time of lip movements). Static signals are irrelevant or even harmful to lip-reading. To eliminate this extraneous information, we calculate the first order difference of phase between two consecutive samples at time $t-1$ and $t$, and denote it as \textit{phase delta}:
\begin{equation}
\Delta \phi_p(t) = \phi_p(t) - \phi_p(t-1).
\end{equation}

Figure~\ref{doppler_pham}(c) shows the phase delta signals of the same sound record. It can be observed that the phase delta signals are approximately zero in the absence of lip movements, which confirms that static signals are almost completely eliminated. In addition, signals with different frequencies fluctuate differently, which shows they experience multipath fading. We also follow the idea in \cite{Kumar2011} and calculate \textit{phase double-delta}, which stands for the second-order difference of phase signals:
\begin{equation}
\Delta \Delta \phi_p(t) = \Delta \phi_p(t) - \Delta \phi_p(t-1).
\end{equation}

All of these phase features can be candidates for our input, and we also try various combinations of these features in our experiments. The details are shown in Table~\ref{input}. Here, the operator $[\cdot,\cdot]$ in the last row represents concatenation operation.

\begin{table}[th]
	\caption{different choices of phase features $x(t)$.}
	\begin{center}
		\begin{tabular}{c|c}
			\hline
			feature name & phase features $x(t)$  \\
			\hline
			\hline
			phase & $\phi_p(t)$  \\
			\hline
			phase delta & $\Delta \phi_p(t)$ \\
			\hline
			phase delta + double-delta & $[\Delta \phi_p(t), \Delta \Delta\phi_p(t))]$  \\
			\hline
		\end{tabular}
		\label{input}
	\end{center}
\end{table}

\textbf{Data augmentation.} In addition, we try to implement data augmentation on the input features $x(t)$. Deep networks usually have a huge demand for the training data. However, it is hard for us to collect sufficient data for lip-reading in practice. In this paper, considering that people speak lip commands at different speeds, we borrow the time-warping technique used in speech recognition \cite{ko2015audio} to enrich training data. Specifically, given a phase signal $x(t)$, we expand or contract the raw signal in the time axis by a factor $\alpha$, thereby generating a new profile $x(\alpha t)$. The data augmentation mechanism can effectively avoid overfitting and improve the robustness against different speeds.

\section {End-to-end Lip-Reading}
\begin{figure}
	\centering
	\includegraphics[width=3in]{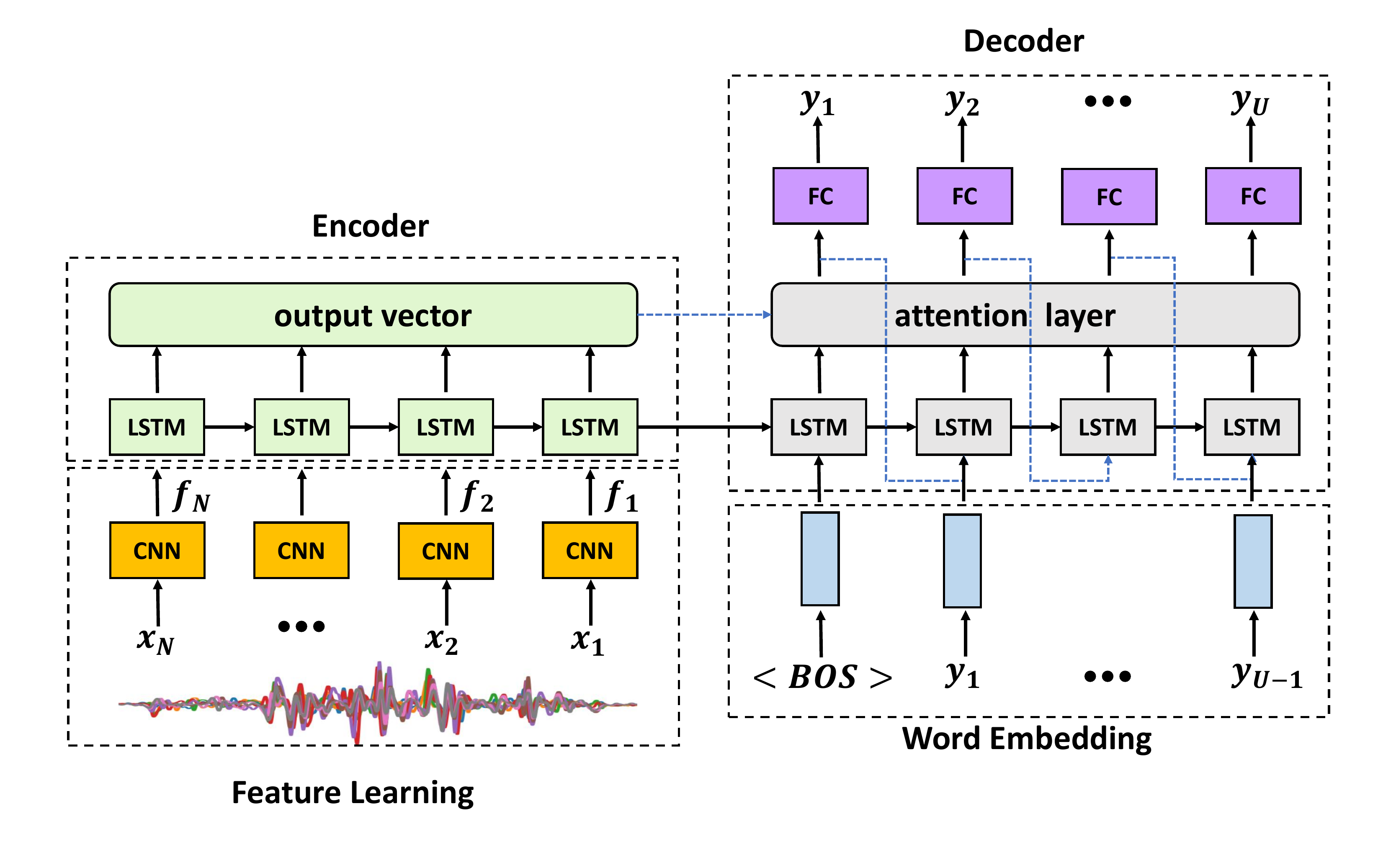}
	\caption{An overview of our end-to-end framework.}
	\label{overview}
\end{figure}

Figure~\ref{overview} illustrates the overall architecture of our framework. The continuous signal stream is firstly segmented into a series of overlapping clips. We employ CNNs to extract features for each clip. As the phase features are temporal signals with different carrier frequencies, we apply convolutions over time and frequency to extract patterns. In particular, assuming there are $N$ clips in total, for the $n$th clip, the input data matrix is $x_{n}^{(T\times{D})}=[x(\tau),x(\tau+1),x(\tau+2),...,x(\tau+T-1)]$, where $D$ is the feature dimension, $\tau$ is the beginning time of this clip and $T$ is the time length in each clip. We input this data matrix into a 3-layer CNN. Pooling and batch normalization (BN) are applied to each layer. After the convolutional layers, fully-connected layers are employed to get the representation vector for this clip.

After feature extraction and processing, we exploit an attention-based encoder-decoder network to achieve end-to-end lip-reading. We break down the sequence learning task into two phases. In the encoding phase, the CNN outputs are projected into a latent space in the form of a fixed size vector, which is later used in the decoding phase to generate sentence labels.

\textbf{Encoder.} In the encoder phase, we implement a 3-layer CNN on $x_n$, and feed the results $f_n$ into a two-layer LSTM, noted as $LSTM_{enc}(\cdot)$, to model temporal changes and output the hidden state $o_n$:
\begin{equation}
f_n = CNN_{\times 3}(x_n),
\end{equation}
\begin{equation}
o_n = LSTM_{enc}(f_n,o_{n+1}).
\end{equation}

Note that the LSTM ingests the inputs in reverse time order, which can shorten long-term dependencies between the beginning of the signal stream and sentence labels, as shown in \cite{sutskever2014sequence}. We denote the final output matrix as $O=[o_N, o_{N-1}, ... ,o_1]$, in which the last column vector $o_1$ corresponds to the latent embedding of the input sequence, and pass it to the decoder.

\textbf{Decoder.} The decoder is also based on a two-layer LSTM, noted as $LSTM_{dec}(\cdot)$. Besides, we utilize the attention mechanism to force the model to learn to focus on specific parts of the input sequence when decoding \cite{bahdanau2014neural}. The key idea of the attention mechanism is to assign a weight $\alpha_u$ for each encoder output $o_n$ at step $u$ of the decoder and generate a context vector $c_u$:
\begin{equation}
\alpha_u = Attention(h_{u-1}, O),
\end{equation} 
\begin{equation}
c_u = O \cdot \alpha_u.
\end{equation}

Then, the hidden state $h_u$ of the decoder at step $u$ can be updated as:
\begin{equation}
y_u, h_u = LSTM_{dec}(h_{u-1}, c_u, g_{u-1}),
\label{yu_cal}
\end{equation}
where $y_u$ is the predicted word label, and $g_{u-1}$ is the word embedding of $y_{u-1}$. The initial $h_0$ is the latent vector $o_1$, and $y_0$ is a special label $<BOS>$ indicating the start of a sentence.  The decoder phase will end when predicting a label $<EOS>$, which indicates the end of the sequence. 

In the training phase, the probability of label $y_u$ at step $u$ can be calculated based on Equation~\ref{yu_cal}. Thus given the signal stream X, the conditional probability of the target sentence $Y$ is:
\begin{equation}
p(Y|X) = \prod_{u} p(y_u).
\end{equation}

We minimize the corresponding cross-entropy loss to update all of the network parameters. In the decoder phase, given the posterior probability distributions of labels at each step, we employ the beam search algorithm to generate the final sequence.

\section{Implementation and Evaluation}
\label{sec:eval}

\begin{figure*}[t]
	\centering
	\subfloat[The WER across volunteers in domain-independent test]{\includegraphics[width=2.2in]{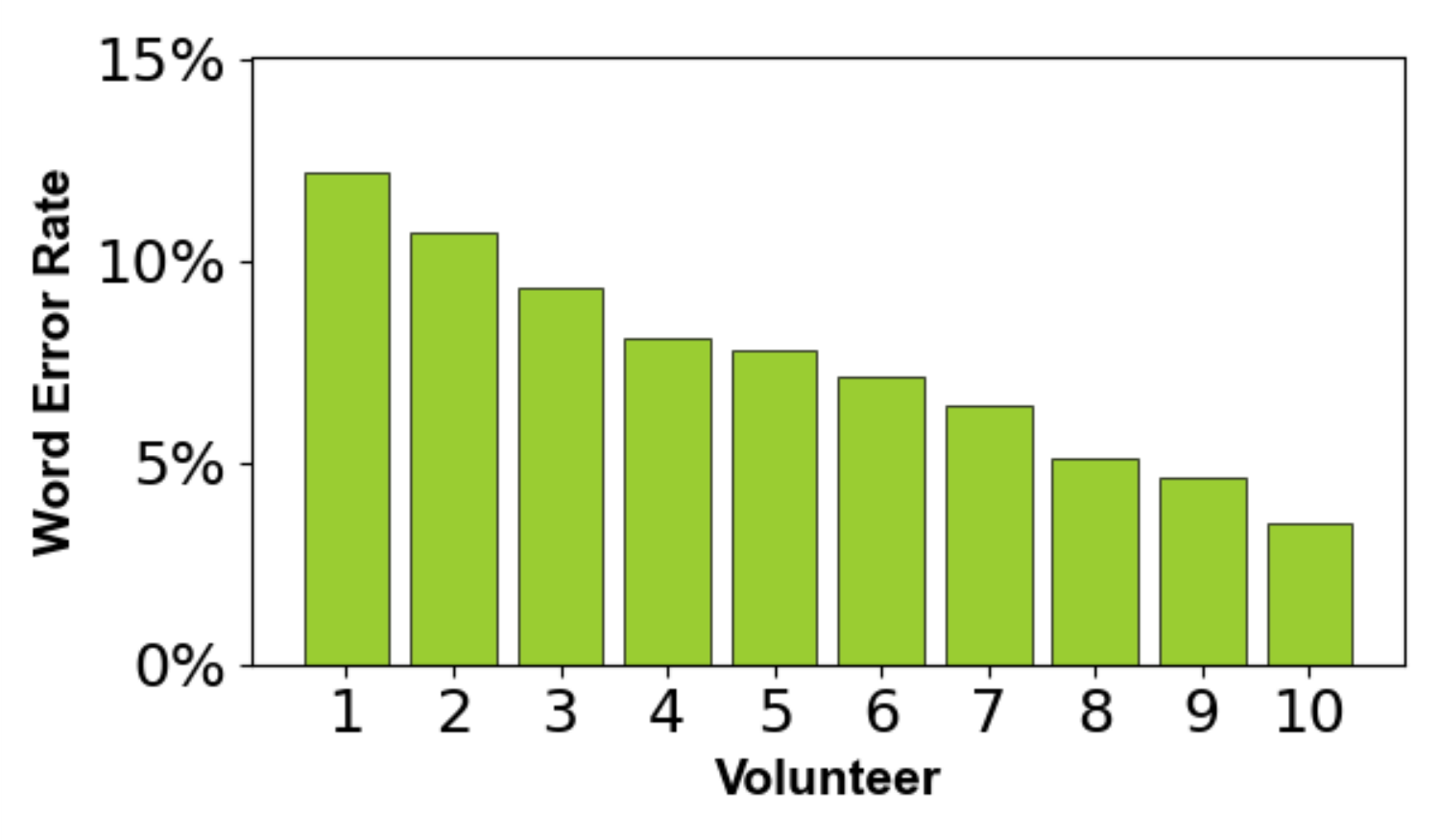}}
	\hspace{5pt}
	\subfloat[Top-10 WER in unseen sentence test]{\includegraphics[width=2.2in]{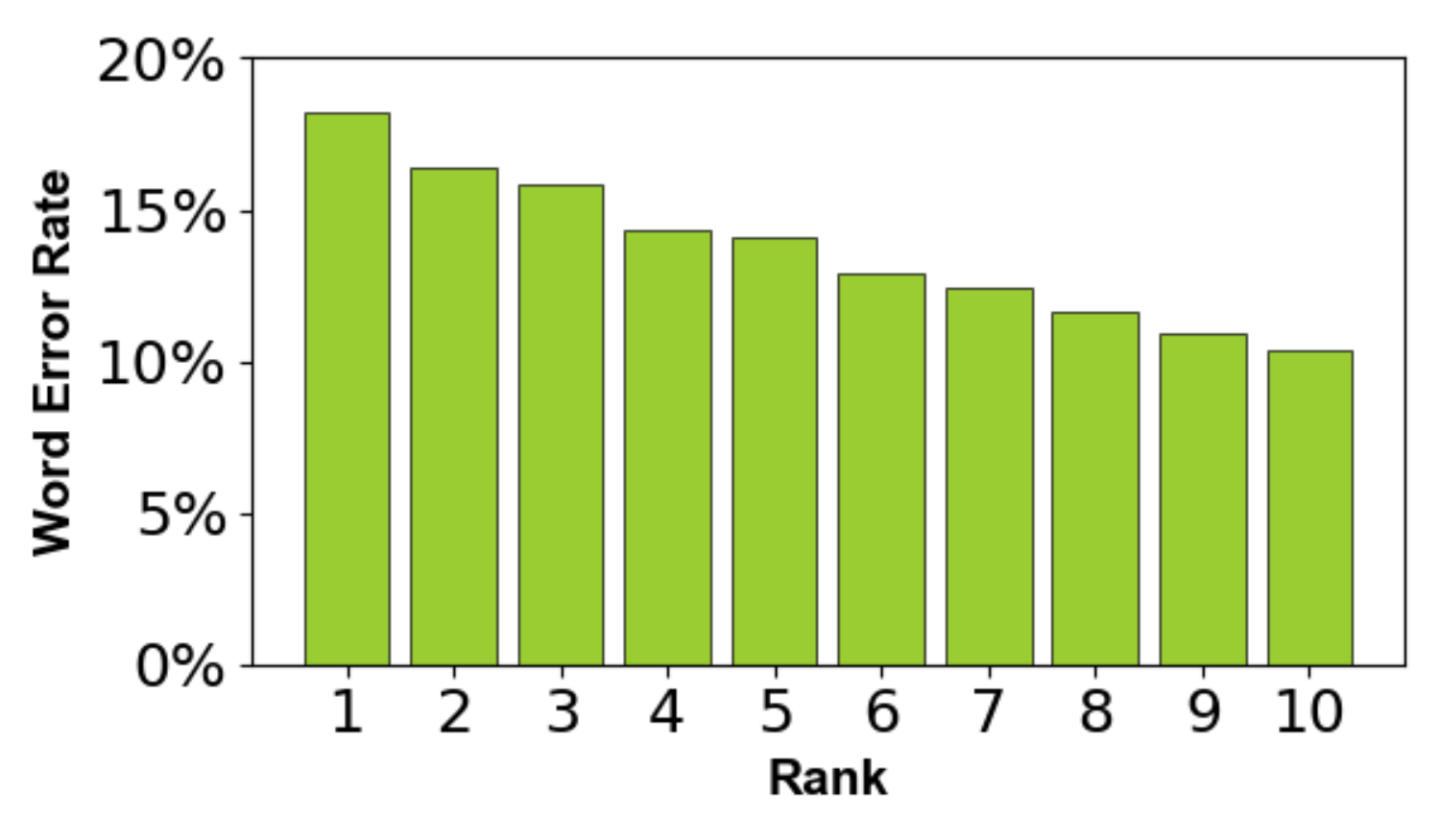}}
	\hspace{5pt}
	\subfloat[Comparison with CTC in different test settings]{\includegraphics[width=2.1in]{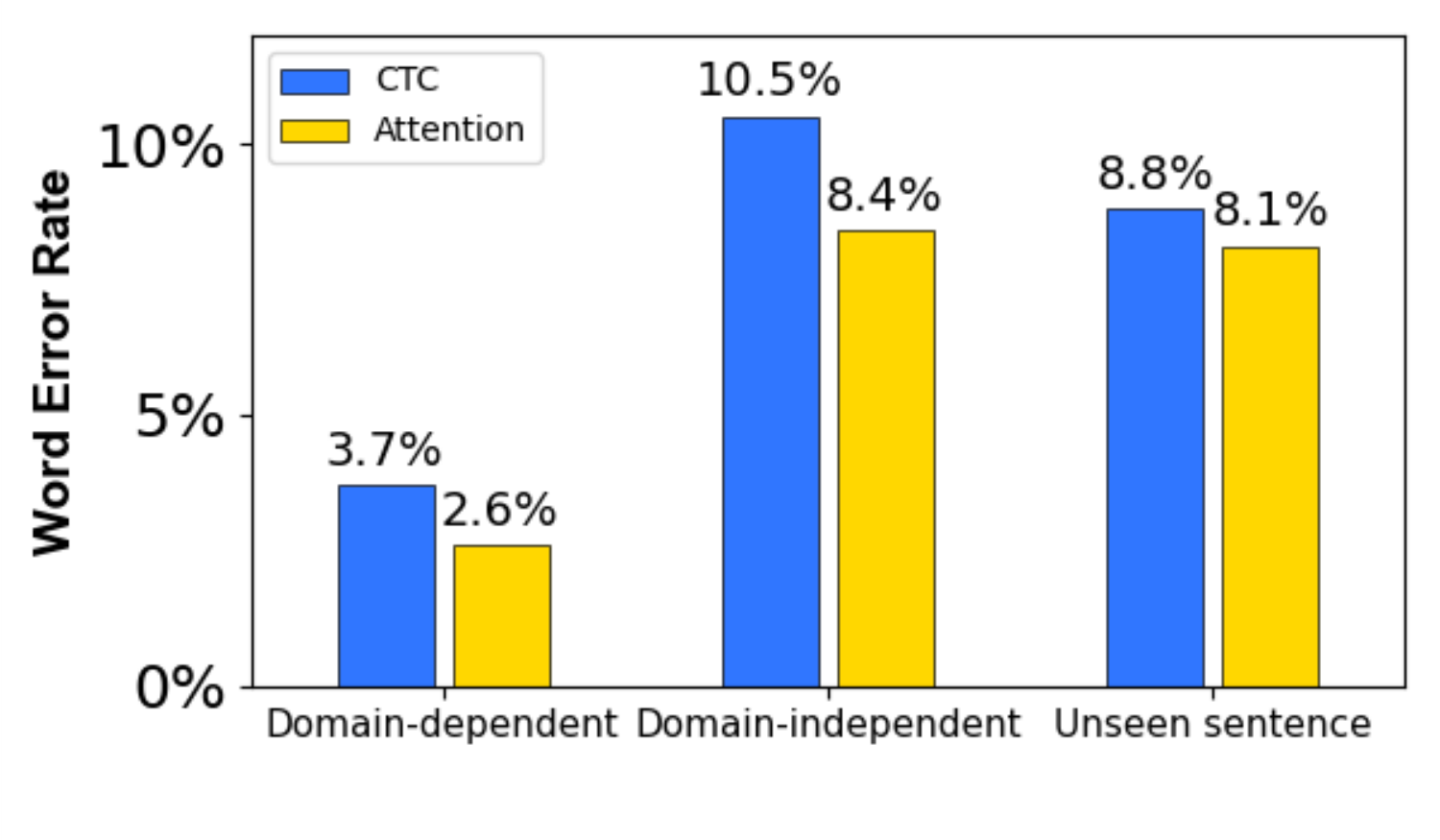}}
	\caption{Experimental results.}
	\label{results}
\end{figure*}


\subsection{Data Collection and Training}
We use one Samsung smartphone to collect lip sentence datasets. All the sentences are Standard Chinese. 10 volunteers are recruited in the data collection. Volunteers perform the experiments at 8 locations (i.e., a laboratory, a meeting room, and a bedroom). Each volunteer performs 5 sessions, and they repeat each sentence 5 times at each session. Different sessions are on different days to ensure data diversity. People can change the distance between the lip and the smartphone according to their habits. We carefully select 54 sentences which are frequently used in voice interfaces of smartphones, e.g., ``What's the weather like tomorrow'' and ``Take a picture on a wide angle'' . A total of 29 words are included in these sentences. In total, we collect 13500 samples. In the training phase, we perform data augmentation for each lip commands with 10 different scaling factor $\alpha$, meaning that the number of samples increases 10 times the original one. The range of $\alpha$ is 0.5$\sim$2.

\subsection{Impact of Signal Processing}

The signal processing pipeline plays a key role in the recognition task. In this paper, we leverage the phase information instead of the coarse-grained Doppler shift to capture lip movements. For all evaluations, we employ the word error rate (WER) as the criterion. Table~\ref{result of sp} lists the WER of different feature choices in the domain-independent test setting.

\begin{table}[h]
	\caption{WER under different signal preprocessing mechanisms}
	\begin{center}
		\begin{tabular}{c|c}
			\hline
			Mechanisms & WER(\%)  \\
			\hline
			\hline
			Doppler shift & 29.4  \\
			\hline
			phase & 32.1  \\
			\hline
			phase delta & 15.6  \\
			\hline
			phase delta + double-delta & 11.2  \\
			\hline
			\textbf{phase delta + double-delta + augmentation} & \textbf{8.4}  \\
			\hline
		\end{tabular}
		\label{result of sp}
	\end{center}
\end{table}

In particular, \textit{Doppler shift, phase, phase delta,} and \textit{phase delta + double-delta} are four types of input features calculated under multiple frequencies. In our experiments, \textit{phase delta + double-delta} gets the lowest WER, outperforming any other features. By the way, we also test \textit{phase delta + double-delta} with only one frequency, but the result (WER 38.2\%) is much worse than using 8 frequency channels. Therefore, the multi-frequency mechanism improves the accuracy by a large margin. Then, we implement data augmentation on \textit{phase delta + double-delta + augmentation} and get the best result in our experiments. We can clearly see that each component in our pipeline boosts the recognition performance to some extent.

\subsection{Evaluation and Performance}
\label{eval_perf}
We evaluate our method by three evaluation strategies: 

\textbf{Domain-dependent test:} Domains in this paper refer to users and environments, both of which have an impact on the phase profiles. Domain-dependent means we ignore the impact of different domains, and randomly divide all the data into training, validation, and testing sets. In the domain-dependent test, we randomly select 70\% of the dataset as training data, 10\% as validation data and 20\% as testing data. The WER of testing data here is \textbf{2.6\%}.

\textbf{Domain-independent test:} We perform leave-one-domain-out cross-validation to validate the capacity of our method to deal with domain diversity. The model hyperparameters are fixed in each test, which are tuned on the validation data of domain-dependent test. We make sure that the training data and the testing data are collected from different users and positions. For the domain-independent test, we present the WERs across volunteers in Figure~\ref{results}(a). From the figure, we get the WERs ranging from 3.5\% to 12.2\%, and the average is \textbf{8.4\%}. The results show that the fine-grained phase profiles as well as the deep learning networks are capable of capturing the key characteristics of these lip sentences, thus generalizing very well across different domains.

\textbf{Unseen sentences test:} We also evaluate the performance in translating unseen sentences (sentences not in the training set). As there is no public and large-scale dataset for acoustic-based lip-reading, the recognition ability for unseen sentences can eliminate the burden to collect all possible sentences. For the unseen sentences test, we perform leave-one-sentence-out validation. We list Top-10 WER in Figure~\ref{results}(b) to show the worst cases. The highest WER is 18.2\%, and the average for all 54 sentences is \textbf{8.1\%}. This is a very impressive result considering that the testing sentences are not included in the training set.

\subsection{Comparison with CTC}
In this paper, we follow the WAS network \cite{chung2017lip} to use the attention-based encoder-decoder framework for sequence modeling. Another popular method used in speech recognition and lip-reading is Connectionist Temporal Classification (CTC) \cite{graves2006connectionist,graves2014towards,assael2016lipnet}. The advantage of the attention model is that it explicitly uses the history of the target label, while CTC assumes the output labels are not conditioned on each other. We make a comparison with CTC by replacing the decoder network with CTC. Figure~\ref{results}(c) presents the result under three evaluation mechanisms. From the figure, we can see that the attention model outperforms CTC notably, especially in the domain-independent conditions. This is mainly due to the ability of the attention to learn internal language models, which is very helpful in the decoding phase.

\section{Conclusion}
In this paper, we propose a non-invasive silent speech recognition method, which uses the inaudible acoustic signals generated by smart devices for lip-reading. We leverage the phase information of the received signals to characterize fine-grained lip movements. And we propose an end-to-end recognition framework which combines the CNN and attention-based encoder-decoder network. We show that the combination of phase delta and double-delta features can get high accuracy on continuous silent speech recognition based on the dataset we collect. The WER under domain-dependent, domain-independent, and unseen sentence tests are 2.6\%, 8.4\%, and 8.1\%, respectively, demonstrating the feasibility and effectiveness of our method. As our method can be seamlessly applied to existing voice-controlled smart devices without any modifications, we believe it can significantly contribute to the advancement of silent voice recognition. Future works include exploring various sequence learning architectures like CTC-attention joint model. We are also interested to combine traditional acoustic signals and the silent signals in speech recognition and voice activity detection tasks.

\section{Acknowledgement}
This paper is supported by National Key Research and Development Program of China under Grant No.2017YFB1401202, No.2018YFB0204400 and No.2018YFB1003500.

\clearpage

\bibliographystyle{IEEEbib}

\bibliography{mybib}

\end{document}